\documentclass[aps,prd,superscriptaddress]{revtex4-2}
\usepackage{graphicx}
\usepackage{subcaption}
\usepackage{natbib}
\usepackage{mathtools}
\usepackage{slashed}
\usepackage{url}
\usepackage[normalem]{ulem}
\usepackage{enumitem}
\usepackage{mathrsfs}
\usepackage{float}
\usepackage{multirow}
\usepackage{xcolor}
\usepackage{booktabs}

\def\be{\begin{eqnarray}}
\def\ee{\end{eqnarray}}
\def\nn{\nonumber}

\usepackage[colorlinks=true, linkcolor=blue, citecolor=violet, urlcolor=teal]{hyperref}
\newcommand{\orcid}[1]{\href{https://orcid.org/#1}{\includegraphics[width=8pt]
		{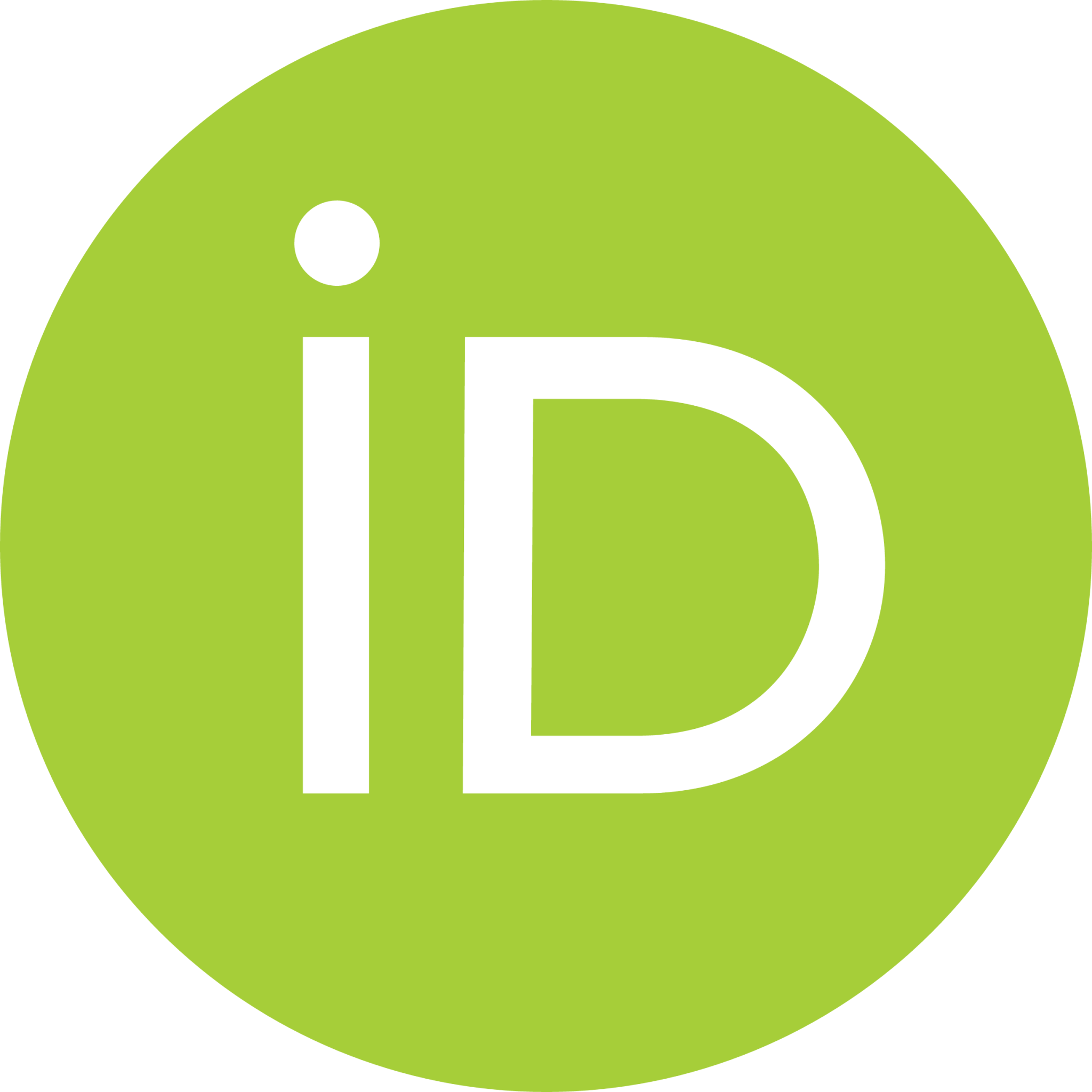}}}
\begin{document}

 \title{Electron-Ion Collision Environment: Distribution of Quark Spin and Orbital Angular Momentum}
\author{Sujit Jana\orcid{https://orcid.org/0009-0000-5469-4229}}
\email{sujitjana011@gmail.com}
\affiliation{Department of Physics, Indian Institute of Technology Bombay, Powai, Mumbai 400076, India}

\author{Ashutosh Dwibedi\orcid{https://orcid.org/0009-0004-1568-2806}}
\email{ashutoshdwibedi92@gmail.com}
\affiliation{Department of Physics, Indian Institute of Technology Bhilai, Kutelabhata, Durg 491002, Chhattisgarh, India}

\author{Vikash Kumar Ojha\orcid{https://orcid.org/0000-0002-0641-4015}}
\email{vko@phy.svnit.ac.in}
\affiliation{Department of Physics, Sardar Vallabhbhai National Institute of Technology, Surat 395007, India}

\author{Sabyasachi Ghosh\orcid{https://orcid.org/0000-0003-1212-824X}}
\email{sabya@iitbhilai.ac.in}
\affiliation{Department of Physics, Indian Institute of Technology Bhilai, Kutelabhata, Durg 491002, Chhattisgarh, India}

\begin{abstract}
The future Electron-Ion Collider (EIC) will enable measurements of the same partonic distributions inside both the proton and the nucleus through electron-proton (eP) and electron-ion (eA) collisions. This capability motivates the present theoretical study of the distributions of quark spin and orbital angular momentum within the proton and the nucleus. To map the eP and eA collision environments, we employ the Nambu-Jona-Lasinio (NJL) model at finite nuclear density to determine the constituent quark masses at zero nuclear density and near the nuclear saturation density. Using these quark mass inputs, we calculate the generalized transverse momentum-dependent parton distributions (GTMDs) associated with quark orbital angular momentum (OAM), spin, and spin-orbit correlations within the light-front dressed quark model. Furthermore, inspired by the nuclear suppression factor widely used in heavy-ion collision experiments, we introduce a set of GTMD ratios between eP and eA collisions. Any deviation of these ratios from unity provides an indirect measure of many-body nuclear density effects arising from non-perturbative quantum chromodynamics (QCD).

\end{abstract}
\maketitle


\section{Introduction}
Protons and neutrons are composite hadrons made of quarks and gluons that are confined by the strong interaction. A central goal of next-generation facilities such as the electron–ion collider (EIC)~\cite{Accardi:2012qut} is to achieve a detailed understanding of the three-dimensional structure of quarks and gluons inside the proton. In particular, these studies aim to clarify how these constituents give rise to the proton's fundamental properties, including its mass, magnetic moment, and spin. The proposed Electron–Ion Collider in China (EicC) is designed to address similar scientific questions~\cite{Anderle:2021wcy}.

From a theoretical perspective, the internal structure of a fast-moving nucleon is commonly described in terms of the one-dimensional parton distribution function (PDF) $f(x)$, which encodes the longitudinal momentum distribution of partons~\cite{Collins:1981uw,Martin:1998sq,Gluck:1994uf,Gluck:1998xa}. However, PDFs provide only a limited, one-dimensional picture of the proton. To obtain a more complete, multidimensional description, one introduces generalized parton distributions (GPDs) $f(x,\xi,t)$~\cite{Ji:1996nm,Diehl:2003ny,Belitsky:2005qn,Goeke:2001tz} and transverse-momentum-dependent distributions (TMDs) $f(x,k_\perp^2)$~\cite{Mulders:1995dh,Barone:2001sp,Bacchetta:2006tn,Brodsky:2002cx,Bacchetta:2017gcc,Kaur:2026wxm}, which incorporate additional spatial and momentum information. At the most comprehensive level in this hierarchy are the GTMDs, which encode the maximal partonic information about the nucleon~\cite{Meissner:2008ay,Meissner:2009ww,Lorce:2013pza}. In the same spirit, different parton distributions inside mesons have also been explored in the literature~\cite{Dwibedi:2025vhr,Dwibedi:2026ozl,Dwibedi:2026rtm,Puhan:2026ppv,P:2026crg,Gautam:2026ohc,Puhan:2025ujg,Tanisha:2025qda,Pandey:2025rqo,Tanisha:2025glu,Yadav:2025txk,Puhan:2025kzz,Puhan:2025ibn,Singh:2024lra,Kaur:2024wze,Acharyya:2024tql}.

GTMDs are defined through the most general off-diagonal quark--quark correlator~\cite{Meissner:2008ay,Meissner:2009ww,Lorce:2013pza}. In recent years, they have attracted considerable attention due to their direct and model-independent connection to partonic OAM and spin--orbit correlations~\cite{bhattacharya2017generalized,hatta2016probing,ji2017hunting,hatta2017gluon,bhattacharya2022signature,hagiwara2017accessing,zhou2016elliptic,lorce2014spin}. Furthermore, GTMDs are expected to become accessible in future experimental studies. The present work aims to probe some selective GTMDs, which are associated with quantities like OAM, spin, and spin-orbit correlation of quarks inside a proton. Their contributions are important for understanding the spin budget of the proton, which is assumed to be coming from a detailed summation of OAM and the spin contribution of quarks and gluons. To evaluate GTMDs and corresponding angular momentum, we use the light-front dressed quark model~\cite{ojha2023quark}; pioneering works in this model are found in Refs. \cite{mukherjee2014quark,Mukkherjee:2015phf,Mukherjee:2015aja,harindranath1999orbital}. Like the spin budget inside the proton, the mass budget also is not realizable if one only adds the masses of two u and one d quarks ($\sim$ 5-10 MeV). However, this mass budget can be understood by using effective QCD models, such as the NJL model. By using the NJL model, one can effectively consider the detailed non-perturbative QCD interaction by designing a temperature ($T$) and chemical potential ($\mu$) or nuclear/baryon density ($\rho_B$) dependent quark mass. First developed for nucleons~\cite{PhysRev.122.345,PhysRev.124.246}, this model was subsequently generalized to quark degrees of freedom to study the thermodynamics of dense quark matter~\cite{Vogl:1991qt,Klevansky:1992qe,HATSUDA1994221,Buballa:2003qv}. In the two-flavor NJL model, the four-fermion contact interaction gives rise to the quark self-energy, which in turn gives rise to a dynamical quark mass (constituent quark mass) different from the current quark mass. In the heavy-ion physics community, this model has been widely used to study the thermodynamic~\cite{Schwarz:1999dj,rai2025towards} and transport~\cite{Marty:2013ita,rai2025towards} properties of the medium. Although it has only quark degrees of freedom, it is widely used as an effective model that captures the medium properties both below (hadronic/nucleonic matter) and above (de-confined quark matter) the phase transition line of the QCD phase diagram~\cite{Ali:2024owl,Ali:2024nrz,Fukushima:2008wg}. At $T=0$ and $\rho_B=0$, the quark mass in the NJL model becomes around 313 MeV, from where the origin of the proton mass $939$ MeV $=3\times 313$ MeV can be well understood. So, one may use the dressed quark with constituent quark mass at $T=0$ and $\rho_B=0$ for eP collision environment and the same at $T=0$ and $\rho_B=\rho_0$ (where $\rho_0=0.16$ fm$^{-3}$ is nuclear saturation density) for eA collision environment. By imposing this innovative bridging between NJL and lightfront QCD dressed quark model, we have theoretically predicted the modification of GTMDs and then evaluated the contribution of the quark to OAM, spin, and spin-orbit correlation, when one goes from eP collision to eA collisions environment. 

The article is organized as follows. In section~\ref{Sec:Form}, we present the theoretical framework used in this work. Specifically, subsection~\ref{Subsec:LFDQM} describes the light-front dressed quark model used to evaluate the GTMDs, while subsection~\ref{Subsec:NJLM} outlines the NJL model and its role in determining the density-dependent constituent quark mass. In section~\ref{Sec:RD}, we present and discuss the numerical results for the GTMDs, together with the corresponding quark OAM, spin, and spin-orbit correlations for proton in vacuum and nuclear environments. Finally, section~\ref{Sec:SC} summarizes the main findings of the present work and presents the concluding remarks.




\section{Formalism}\label{Sec:Form}
Here, we first address the formalism of GTMD briefly by using the light front dressed quark model~\cite{ojha2023quark} having constituent quark mass $m^*$. Then, we briefly describe the NJL model formalism, which imposes the non-perturbative QCD environment in eP and eA collisions using the quark mass in vacuum and at nuclear saturation density.

\subsection{Light front Dressed quark model}\label{Subsec:LFDQM}
We opt for a light-front coordinate system denoted as $(z^+, z^-, z_\perp)$, where the light-front time variable ($z^+$) and the light-front longitudinal space variable ($z^-$) are defined as $z^{\pm} = z^0 \pm z^3 $~\cite{harindranath1996introduction,zhang1994light}. Our focus involves a system comprising a quark dressed with a gluon, subjected to probing by a virtual photon. The total squared momentum transferred from the virtual photon to the target state is denoted as $t=(p-p^\prime)^2=\Delta^2$. The longitudinal momentum transferred to the target is expressed through the skewness variable $\xi$, defined as $\xi = \Delta^+/2P^+$. We adopt a symmetric frame and parameterize the initial and final four-momentum of the target state as~\cite{brodsky2001light}
\be
    p &=&\Bigg((1+\xi)P^+,\Delta_\perp/2,\frac{m^{\ast 2}+\Delta_\perp^2/4}{(1+\xi)P^+}\Bigg) ;\\
    p'&=&\Bigg((1-\xi)P^+,-\Delta_\perp/2,\frac{m^{\ast 2}+\Delta_\perp^2/4}{(1-\xi)P^+}\Bigg),
\ee
where $P=(p + p^\prime)/2$ is the average momentum of the target state, and the  momentum transferred to the target state is 
\be
    \Delta = p-p' = \Bigg(2\xi P^+, \Delta_\perp, \frac{t+\Delta_\perp^2}{2\xi P^+} \Bigg),
\ee
with
\be
    t=-\frac{4\xi^2 m^{\ast 2} + \Delta_\perp^2}{1-\xi^2}.
    \ee
Here, $m^{\ast}$ is the in-medium mass of the dressed quark obtained in the NJL model that is discussed in the next subsection. The quark carries the average longitudinal momentum fraction $x=k^+/P^+$ of the target state, and the quark four-momentum is
\be 
k\equiv \Big(x P^+, k_{\perp }, k^- \Big)\label{mom_k}.
\ee


To study the GTMDs, we are considering a dressed quark as the target state. A dressed quark can be considered as a bound-state system with spin $1/2$, consisting of a bare quark and a gluon. The model has been used previously to study the behavior of quark and gluon in a bound-state system. A dressed quark state with momentum $p$ and spin $\sigma$ can be expanded in the Fock space as\cite{zhang1993light,mukherjee2014quark,ojha2023quark} 
\be \label{fockse}
  \Big{| }p^{+},p_{\perp},\sigma  \Big{\rangle} = \Phi^{\sigma}(p) b^{\dagger}_{\sigma}(p)
 | 0 \rangle +
 \sum_{\sigma_1 \sigma_2} \int [dp_1]
 \int [dp_2] \sqrt{16 \pi^3 p^{+}}
 \delta^3(p-p_1-p_2) \nn \\ \Phi^{\sigma}_{\sigma_1 \sigma_2}(p;p_1,p_2) 
b^{\dagger}_{\sigma_1}(p_1) 
 a^{\dagger}_{\sigma_2}(p_2)  | 0 \rangle;
\ee
where, we truncated the series expansion up to two-particle state. Here, $[dp] =  \frac{dp^{+}d^{2}p_{\perp}}{ \sqrt{16 \pi^3 p^{+}}}$, $\Phi^\sigma(p)$ is the single-particle wavefuction with momentum $p$ and spin $\sigma$. $\Phi^{\sigma}_{\sigma_1 \sigma_2}(p;p_1,p_2)$ is the two-particle light front wave function and can be expressed in the boost-invariant form using the relation  $\Psi^{\sigma}_{\sigma_1
\sigma_2}(x, q_\perp) =   
\Phi^{\sigma}_{\sigma_1 \sigma_2}
\sqrt{P^+}$.
The momentum variables $(x_i, q_{i \perp})$ are Jacobi momenta defined as 
\be \label{jacmom}
p_i^+= x_i p^+, ~~~~~~~~~~q_{i \perp}= k_{i \perp}+x_i p_\perp,
\ee
and satisfies the following constraints 
\begin{equation*}
    \sum_i x_i=1, ~~~~~~~~~\sum_i q_{i\perp}=0
\end{equation*}
Therefore, the boost invariant two-particle LFWFs defined as \cite{harindranath1999orbital,more2017quark}

\be \label{tpag}
\Psi^{\sigma a}_{\sigma_1 \sigma_2}(x,q_{\perp}) = 
\frac{1}{\Big[    m^{\ast 2} - \frac{m^{\ast 2} + (q_{\perp})^2 }{x} - \frac{(q_{\perp})^2}{1-x} \Big]}
\frac{g}{\sqrt{2(2\pi)^3}} T^a \chi^{\dagger}_{\sigma_1} \frac{1}{\sqrt{1-x}}
\nn \\ \Big[ 
-2\frac{q_{\perp}}{1-x}   -  \frac{(\sigma_{\perp}.q_{\perp})\sigma_{\perp}}{x}
+\frac{im^{\ast}\sigma_{\perp}(1-x)}{x}\Big]
\chi_\sigma (\epsilon_{\perp \sigma_2})^{*}.
\ee
Here, $T^a$ is the $SU(3)$ color matrices, $m$ is the mass of the dressed quark, $\chi$ is the 2-component spinor, and $\epsilon_\perp$ is the polarization vector of the gluon, as taken from the Refs. \cite{diehl2003generalized,mukherjee2013generalized}. We choose initial and final Jocbi momenta for the target state as $(y,q_\perp)$ and $(x',q'_\perp)$, respectively. Using the above wave functions, this model can investigate non-perturbative distributions for quarks, e.g., GPDs, TMDs, GTMDs, etc.

In light-front gauge with $z^+=0$, the quark-quark correlator $W_{\lambda,\lambda'}^\Gamma(x,\xi,\Delta_\perp,k_\perp;S)$ is defined via the off-diagonal matrix element of the bi-local quark field, as given below~\cite{mukherjee2014quark}:
\begin{align}\label{qqc}
    W_{\lambda,\lambda'}^{[\Gamma]}(x,\xi,\Delta_\perp,k_\perp)=&\frac{1}{2}\int\frac{dz^-}{2\pi}\frac{d^2z_\perp}{(2\pi)^2}e^{ip.z}\Big<p',\lambda'\Big|\Bar{\psi}(-\frac{z}{2})\mathcal{W}_{[-\frac{z}{2},\frac{z}{2}]}\Gamma\psi(\frac{z}{2})\Big|p,\lambda\Big>\Bigg|_{z^+=0},
\end{align}
Here, $|p,\lambda\rangle$ and $|p',\lambda'\rangle$ denote the initial and final states of the dressed quark system, respectively. The Wilson line $\mathcal{W}_{[-\frac{z}{2},\frac{z}{2}]}$ acts as a gauge link connecting the quark fields $\psi(\frac{z}{2})$ and $\Bar{\psi}(-\frac{z}{2})$ located at distinct spacetime points.
Using the dressed quark state and the quark field's particle sector $\psi(\pm\frac{z}{2})$, the correlator can be expressed in terms of the overlap representation of light-front wave functions. The quark-quark correlators corresponding to unpolarized, longitudinally polarized 
quark states are given by~\cite{mukherjee2014quark,more2017quark}:
\begin{align}
W^{(\gamma^{+})}_{\lambda\lambda'}(x,\xi,k_\perp,\Delta_\perp)&=\displaystyle\sum_{{\sigma_{1}},{\sigma_{2}},{\lambda_1}}\Psi_{\lambda_{1}\sigma_{2}}^{{*}{\lambda'}}{(x',q'_{\perp})}\;\chi^\dagger_{\lambda_1}\chi_{\sigma_1}\;{{\Psi_{{\sigma_{1}}{\sigma_{2}}}^{{\lambda}}}{(y,q_{\perp})}},\label{qqcf1}\\
W^{(\gamma^{+}\gamma_5)}_{\lambda\lambda'}(x,\xi,k_\perp,\Delta_\perp)&=\displaystyle\sum_{{\sigma_{1}},{\sigma_{2}},{\lambda_1}}\Psi_{\lambda_{1}\sigma_{2}}^{{*}{\lambda'}}{(x',q'_{\perp})}\;\chi^\dagger_{\lambda_1}\sigma_3\chi_{\sigma_1}\;{{\Psi_{{\sigma_{1}}{\sigma_{2}}}^{{\lambda}}}{(y,q_{\perp})}},
\label{qqcf2}
\end{align}
where the initial (final) struck quark carries a longitudinal momentum fraction $y(x')$ and transverse momentum $q_\perp(q_\perp')$, with the chosen kinematics parametrized as follows~\cite{ojha2023quark}:
\begin{align}
    x'&=\frac{x-\xi}{1-\xi} ~~,~~~
    q'_{\perp}=k_\perp-\frac{(1-x)}{(1-\xi)}\frac{\Delta_\perp}{2},~~y=\frac{x+\xi}{1+\xi} ~~,~~~
    q_{\perp}=k_\perp+\frac{(1-x)}{(1+\xi)}\frac{\Delta_\perp}{2},\nn
\end{align}
where $x$ denotes the momentum fraction associated with the average quark momentum $k$. We compute the quark-quark correlators, Eqs. (\ref{qqcf1}-\ref{qqcf2}), for all spin combinations $(\lambda\lambda'=\uparrow\uparrow,\uparrow\downarrow,\downarrow\uparrow,\downarrow\downarrow)$ corresponding to each polarization state of the quark. The bilinear decomposition of the quark-quark correlator in Eq. \ref{qqc}, at leading twist, is presented for the cases of unpolarized, longitudinally polarized
quarks as follows~\cite{meissner2009generalized}:
\begin{align}\label{unpara}
W^{[\gamma^+]}_{\lambda,\lambda'}=&\frac{1}{2m^{\ast}}\Bar{u}(p',\lambda')\Big[F_{1,1}-\frac{i\sigma^{i+}k_{i\perp}}{P^+}F_{1,2}-\frac{i\sigma^{i+}\Delta_{i\perp}}{P^+}F_{1,3}+\frac{i\sigma^{ij}k_{i\perp}\Delta_{j\perp}}{m^{\ast 2}}F_{1,4}\Big]u(p,\lambda),
\end{align}
\begin{align}
W^{[\gamma^{+}\gamma_5]}_{\lambda,\lambda'}=&\frac{1}{2m^{\ast}}\Bar{u}(p',\lambda')\Big[\frac{-i\epsilon^{ij}_{\perp}k_{i\perp}\Delta_{j\perp}}{m^{\ast 2}}G_{1,1}-\frac{i\sigma^{i+}\gamma_5 k_{i\perp}}{P^+}G_{1,2}-\frac{i\sigma^{i+}\gamma_5 \Delta_{i\perp}}{P^+}G_{1,3}+i\sigma^{+-}\gamma_5 G_{1,4}\Big]u(p,\lambda),
\label{tpara}
\end{align}
where $u_\uparrow(p)$ and $u_\downarrow(p)$ are the Dirac spinors defined as~\cite{harindranath1996introduction}
\begin{align}
    u_\uparrow(p)=\frac{1}{\sqrt{2p^+}}\begin{pmatrix}
    p^++m^{\ast}\\
    p^1+ip^2\\
    p^+-m^{\ast}\\
    p^1+ip^2
    \end{pmatrix},~~~
    u_\downarrow(p)=\frac{1}{\sqrt{2p^+}}\begin{pmatrix}
    -p^1+ip^2\\
    p^++m^{\ast}\\
    p^1-ip^2\\
    -p^++m^{\ast}
    \end{pmatrix}.
\end{align}
We obtained the various bilinear decompositions of the quark-quark correlator, Eqs. (\ref{unpara}-\ref{tpara}), for all spin combinations $(\lambda\lambda'=\uparrow\uparrow,\uparrow\downarrow,\downarrow\uparrow,\downarrow\downarrow)$ corresponding to each quark polarization state. By comparing Eq. (\ref{qqcf1}-\ref{qqcf2}) with Eq. (\ref{unpara}-\ref{tpara}), we extracted the 8 GTMDs \cite{ojha2023quark}. Among them, the present work focuses only on 3 GTMDs - $F_{1,4}$, $G_{1,1}$, and $G_{1,4}$ for eP and eA collisions environments. The reason for choosing only these 3 GTMDs is that their appropriate integrated values are associated with quark OAM, spin, and spin-orbit correlation, which can be linked with the Proton's spin budget. 
The analytical expressions of these 3 GTMDs in the dressed quark model are given as follows:
\begin{align}
    F_{1,4}=&\frac{\alpha}{(1-x)}\Big[2m^{\ast 2}(1+x)(1-\xi^2)\Big],\label{F14}\\
           G_{1,1}=&\frac{\alpha}{(1-x)}\Big[-2m^{\ast 2}(1+x)(1-\xi^2)\Big]\label{G11},\\
           G_{1,4}=&\frac{\alpha(1-\xi^2)}{4(1-x)^3}\Big[(1+x^2-2\xi^2)(4(1-\xi^2)k^2_\perp+(1-x)(4\xi k_\perp-(1-x)\Delta_\perp)\cdot\Delta_\perp)\nonumber\\
           &-4m^{\ast 2}(1-x)^4\Big].\label{G14}
\end{align}
       
The functions $D(k_\perp,x)$ and $\alpha(x,\xi,k_\perp^2,\Delta_\perp^2,k_\perp.\Delta_\perp)$ are defined as
\begin{align}
    D(k_\perp,x)=m^{\ast 2}-\frac{m^{\ast 2}+(k_{\perp})^{2}}{x}-\frac{(k_{\perp})^{2}}{1-x}\\
    \text{and},~ \alpha(x,\xi,k_\perp^2,\Delta_\perp^2,k_\perp.\Delta_\perp)=\frac{N}{D(q_\perp,y)D^*(q_\perp',x')(x^2-\xi^2)}.
\end{align}
Here $N=\frac{g^2C_f}{2(2\pi)^3}$, where $g$ denotes the strong coupling constant and $C_f$ the color factor. We supress the notation by denoting $\alpha(x,\xi,k_\perp^2,\Delta_\perp^2,k_\perp.\Delta_\perp)$ as $\alpha$.


The GTMD $F_{1,4}$ and $G_{1,4}$ contribute to the connection between the quark's OAM and the spin of the system, given by~\cite{ojha2023quark}:
\begin{align}
    l^q_z=&-\int dxd^2k_\perp\frac{k_\perp^2}{m^{\ast 2}}F_{1,4}\label{lqz}\\
    S^q=&\int dx d^2k_\perp G_{1,4}\label{sq}.
\end{align}

If $l^q_z > 0$, quark OAM tends to align along the spin of the system and for $l^q_z < 0$ they are tending to anti-align. The numerical integration was performed over the $k_\perp$. Ideally, the upper limit of $k_\perp$ integration should be infinite. But according to the standard practice of numerical integration, we took an upper cutoff and chose the upper and lower limit of $k_ \perp$ integration as $Q$ and $0$, respectively. In our model, $l^q_z = -0.126653$ for $Q=5$ GeV, where $Q$ is the large scale involved in the process, indicating that the quark OAM is antialigned with the spin of the dressed quark system.

Furthermore, the GTMD $G_{1,1}$ is linked to the correlation between the quark spin and OAM
\begin{align}
    C^q_z=\int dxd^2k_\perp\frac{k_\perp^2}{m^{\ast 2}}G_{1,1}\label{cqz}.
\end{align}

\subsection{NJL Model}\label{Subsec:NJLM}
Here, we discuss the NJL framework, which can effectively map the dressed quark mass in eP and eA collision environments.
Let us start from the Lagrangian density of the two-flavor NJL model~\cite{Vogl:1991qt,Klevansky:1992qe,HATSUDA1994221,Buballa:2003qv}, 
\be
\mathcal{L}=\bar{\psi}(i\slashed{\partial}-{\mathbf m})\psi +G\left[(\bar{\psi}\psi)^{2}+(\bar{\psi}i\gamma_{5}\vec{\sigma}\psi)^{2}\right]~,\label{AD1}
\ee
where $\psi\equiv\begin{pmatrix}
\psi_{u} \\
\psi_{d}
\end{pmatrix}$ and $\mathbf{m}\equiv\begin{pmatrix}
m_{u} \\
m_{d}
\end{pmatrix}$ are the quark fields and bare mass matrix for different flavors, respectively. In this work, we assume the same mass for quark flavors, \textit{i.e.}, $m_{u}=m_{d}=m$. $\vec{\sigma}$ and $G$ respectively denote the Pauli matrices and the coupling constant. The quark-quark interaction through effective 4-point contact interactions is captured by the second term in the square bracket of the NJL Lagrangian~\eqref{AD1}. The 4-point contact interactions lead to quark self-energy, and the constituent quark mass $m^{*}$ is represented \textit{via} the following gap equation, 
\be
&& m^{*} =m+2iG\int_{\Lambda} \frac{d^{4}p}{(2\pi)^{4}}~TrS(p)~,\label{AD2}
\ee
where $S(p)$ is the dressed propagator, which is given by $S(p)=\frac{1}{\slashed{p}-m^{*}+i\epsilon}$. The trace in the above equation should be taken over color, flavor, and Dirac spaces. The NJL Lagrangian with the contact interaction terms is non-renormalizable, and one introduces a cut-off energy scale $\Lambda$ to regularize the divergent integrals, which show up in the calculations. For different regularization schemes practiced in the NJL model, readers can refer to \cite{Klevansky:1992qe}. Eq.~\eqref{AD2} illustrates the spontaneous breaking of chiral symmetry in which a light current quark mass $m$ turns into a massive constituent quark mass $m^{*}$. The three model parameters--$m$, $G$, and $\Lambda$ are suitably chosen to fit the experimental values of pion mass $m_{\pi}=135$ MeV and pion decay constant $f_{\pi}=92.4$ MeV in vacuum. We have chosen the following values for the parameters in the present paper: $m=0.0027$ GeV, $G=1.95$ GeV$^{-2}$, and $\Lambda=0.95$ GeV. 

For finite nuclear or baryon density $\rho_{B}$, we can replace the dressed propagator $S(p)$ by an in-medium propagator with finite density $\rho_{B}$ (or equivalently quark chemical potential $\mu$). In this finite nuclear density scenario, considering the zero temperature limit we get,
\be
m^{*}=m+\frac{6Gm^{*}}{\pi^{2}}\left[\Lambda\sqrt{\Lambda^{2}+m^{*2}}-\mu\sqrt{\mu^{2}-m^{*2}}-m^{*2}\ln \frac{\Lambda+\sqrt{\Lambda^{2}+m^{*2}}}{\mu+\sqrt{\mu^{2}-m^{*2}}}\right]~.\label{AD4}
\ee
Eq.~\eqref{AD4} can be solved to get the constituent quark mass $m^{*}$ as a function of quark chemical potential $\mu$. One can also readily obtain the constituent quark mass $m^{*}$ as a function of baryon density $\rho_{B}$ by using $\mu=\sqrt{\left(\frac{3\pi^{2}\rho_{B}}{2}\right)^{2/3}+m^{*2}}$ . We will use the constituent quark mass $m^*$ at $\rho_B=0$ and $\rho_B=\rho_0$ for mapping the non-perturbative QCD domain of eP and eA collision environments.

\section{Results and Discussion}\label{Sec:RD}
\begin{figure}[htp!]
\centering
\begin{subfigure}[b]{0.4\textwidth}
\centering
\includegraphics[width=\textwidth]{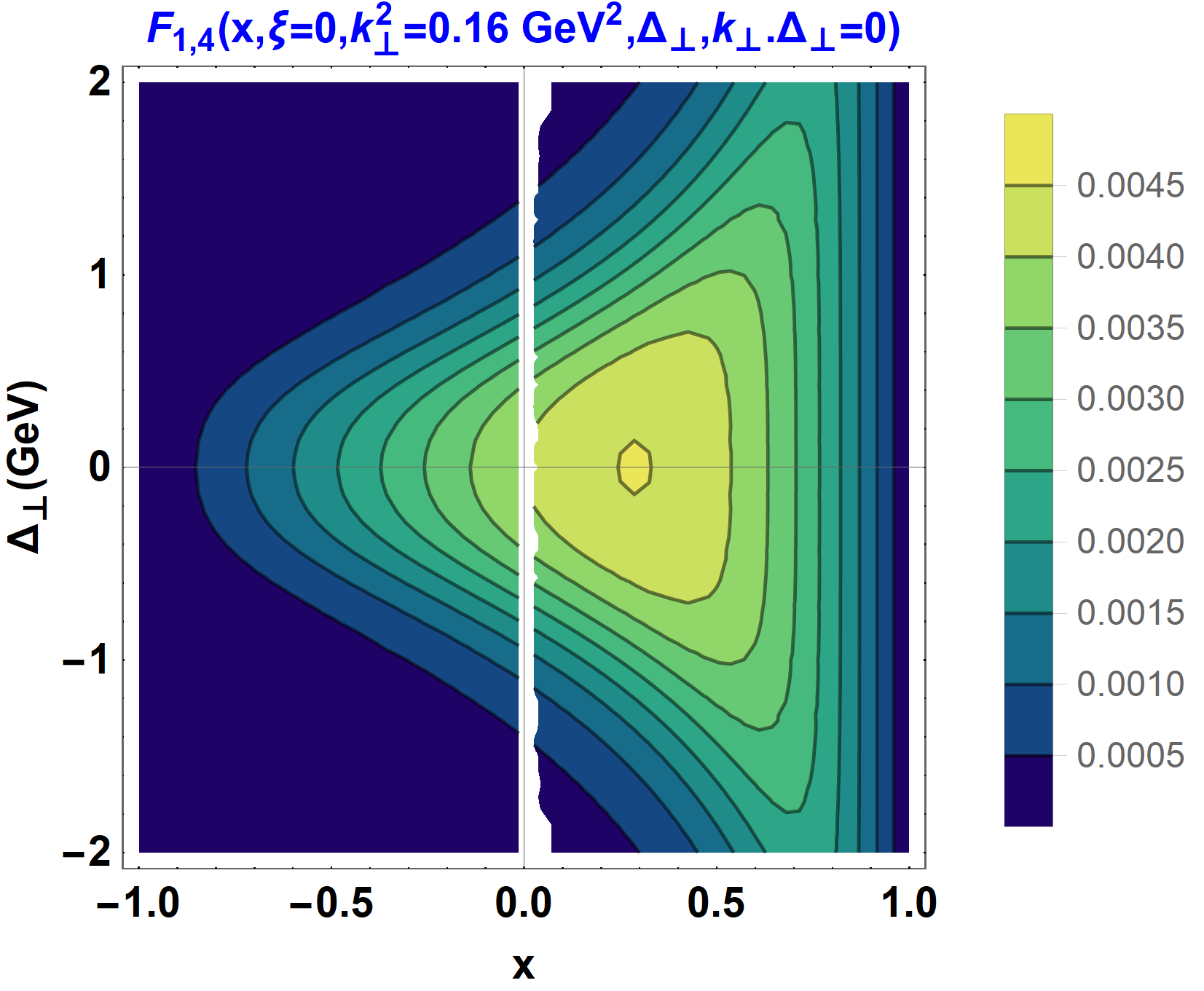}
\caption{$\rho_{B}=0$}
\end{subfigure}
\centering
\begin{subfigure}[b]{0.4\textwidth}
\centering
\includegraphics[width=\textwidth]{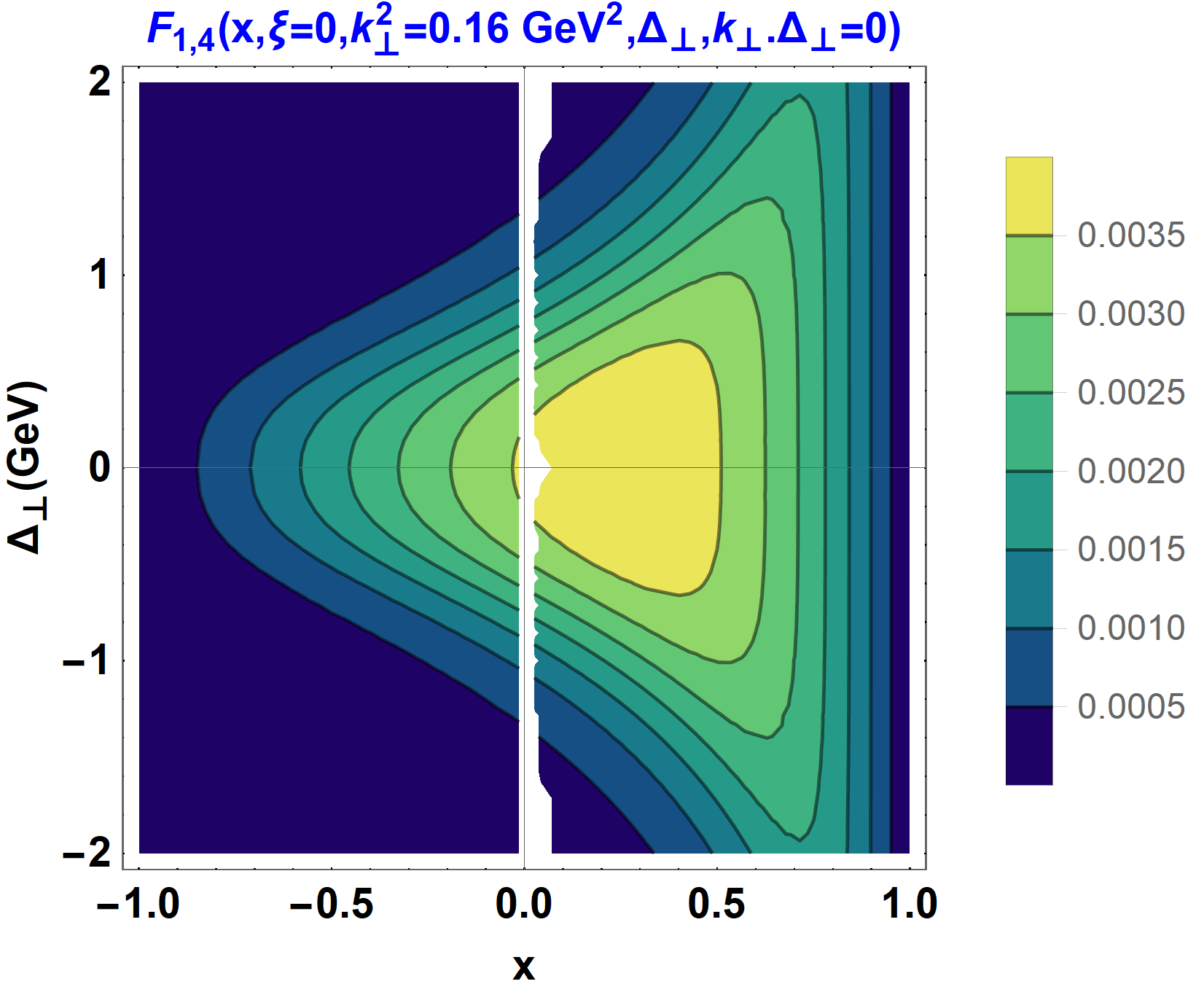}
\caption{$\rho_{B}=\rho_0$}
\end{subfigure}
\hfill
\begin{subfigure}[b]{0.4\textwidth}
\centering
\includegraphics[width=\textwidth]{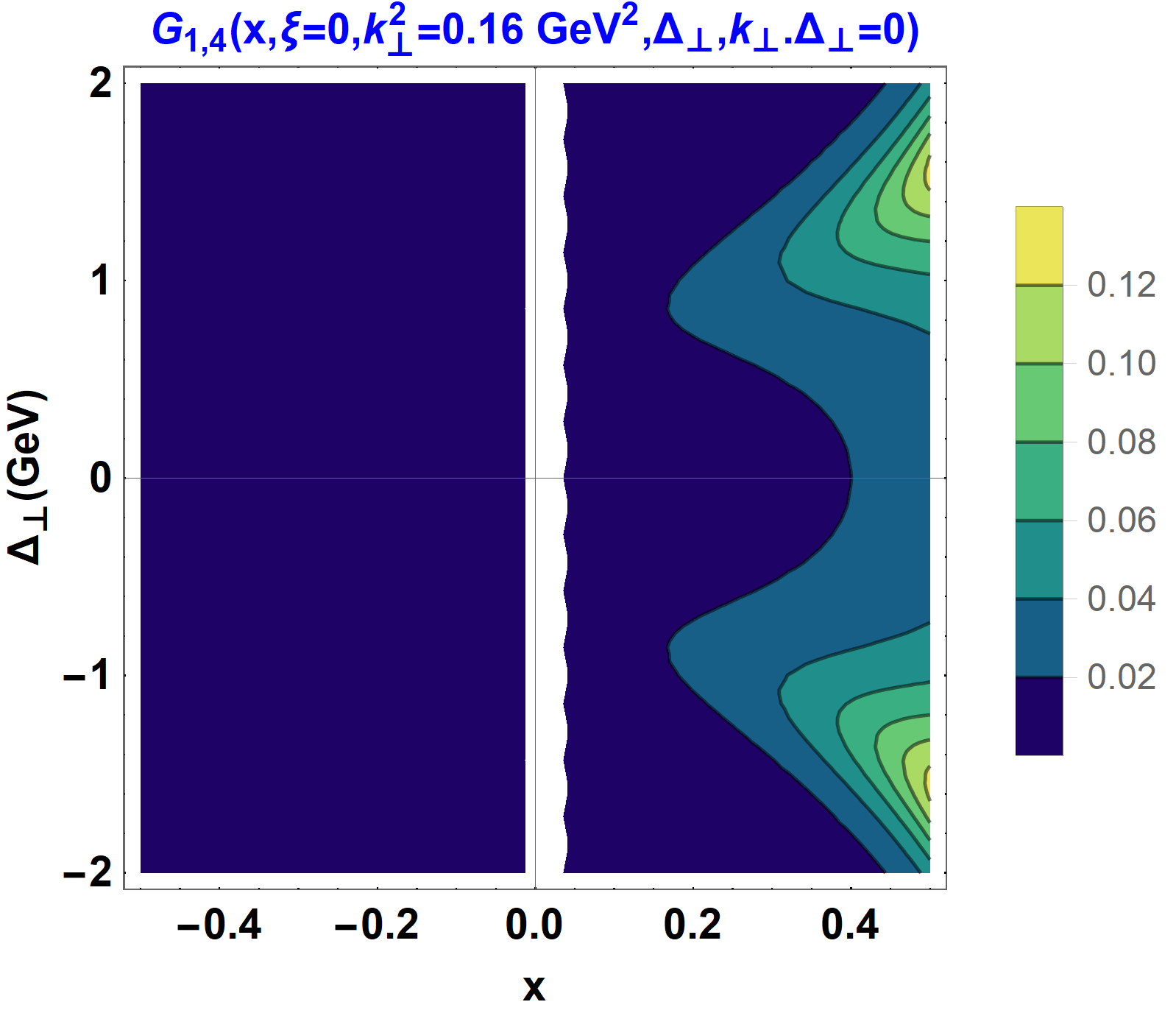}
\caption{$\rho_{B}=0$}
\end{subfigure}
\centering
\begin{subfigure}[b]{0.4\textwidth}
\centering
\includegraphics[width=\textwidth]{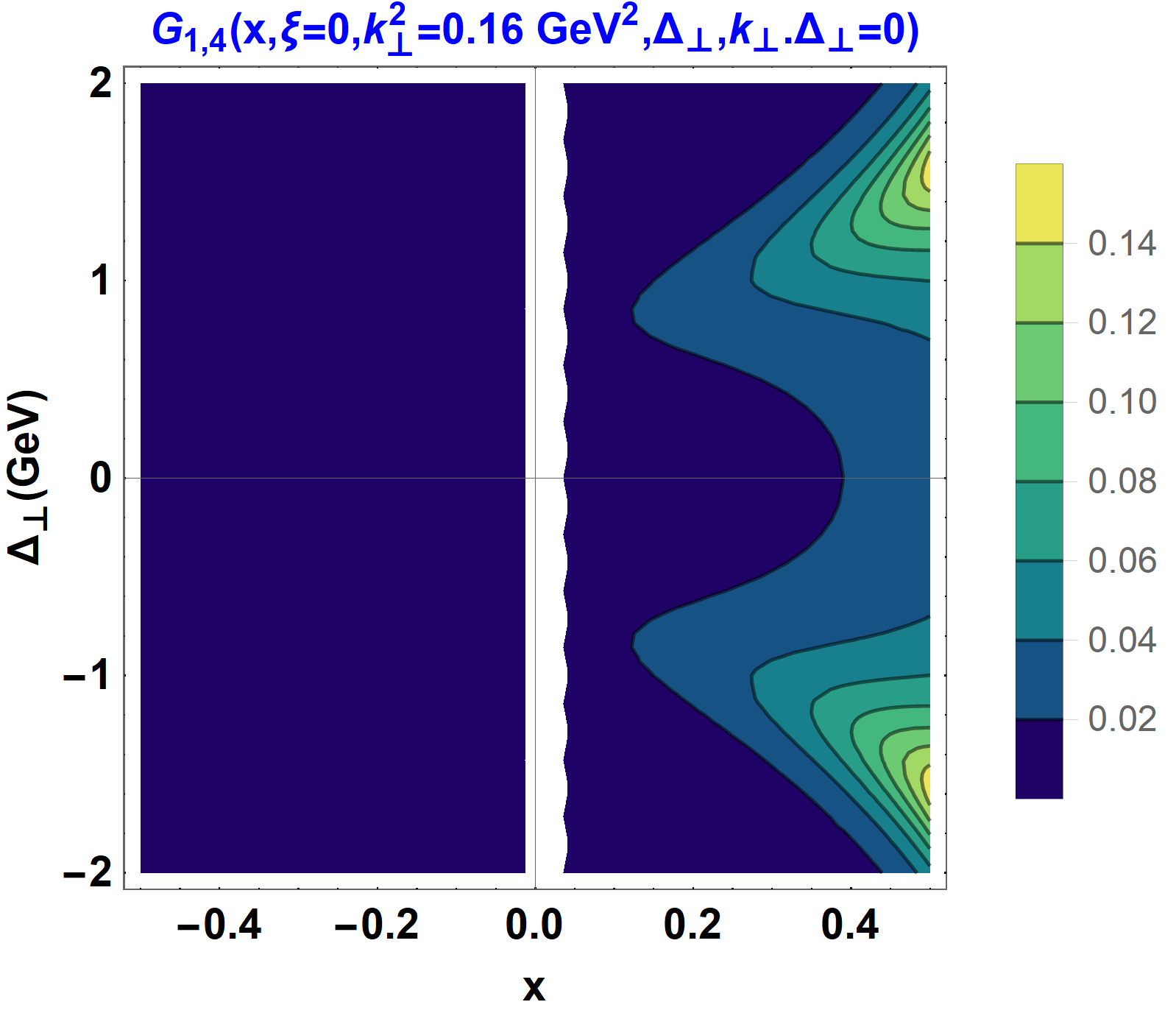}
\caption{$\rho_{B}=\rho_0$}
\end{subfigure}
\hfill
\begin{subfigure}[b]{0.4\textwidth}
\centering
\includegraphics[width=\textwidth]{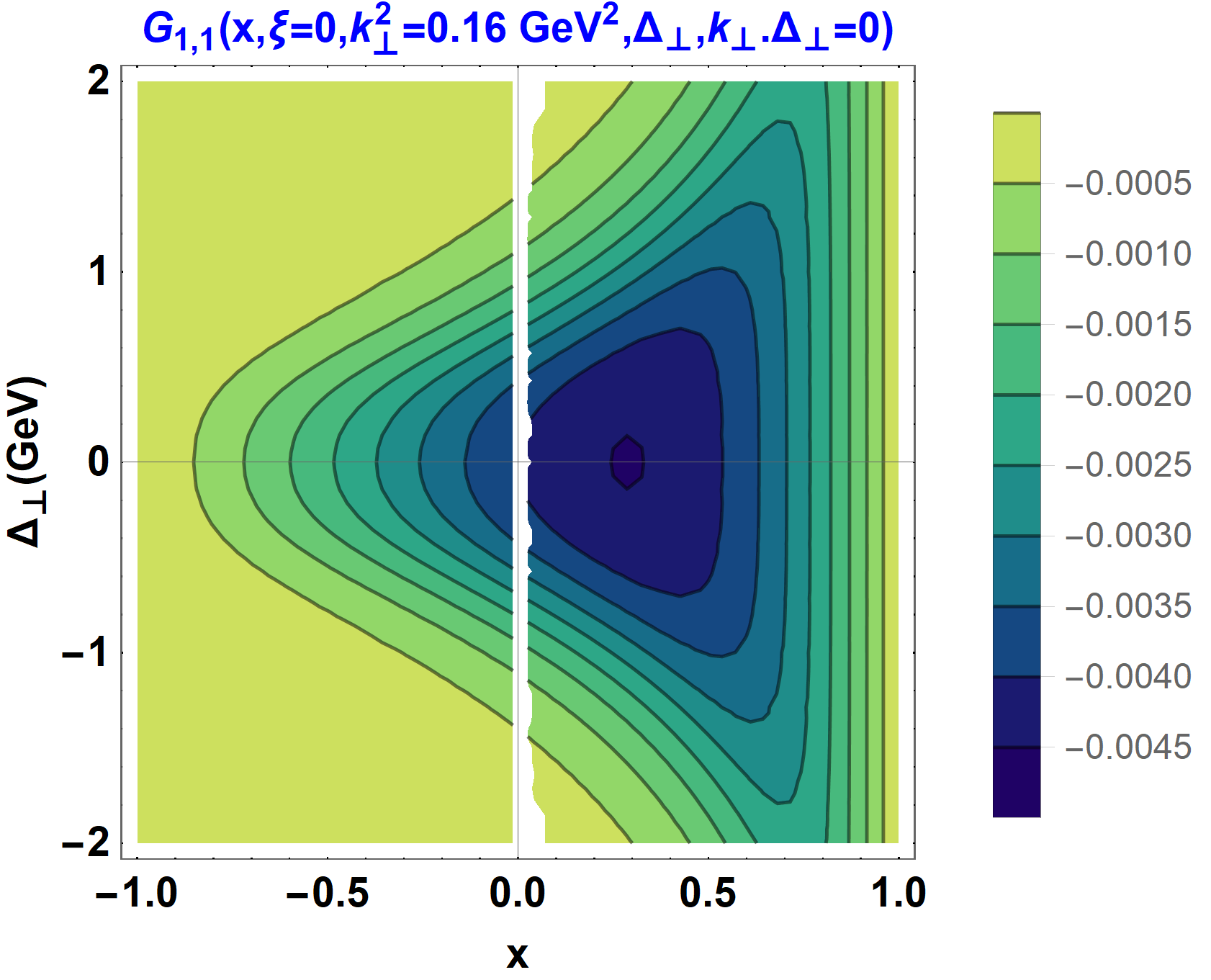}
\caption{$\rho_{B}=0$}
\end{subfigure}
\centering
\begin{subfigure}[b]{0.4\textwidth}
\centering
\includegraphics[width=\textwidth]{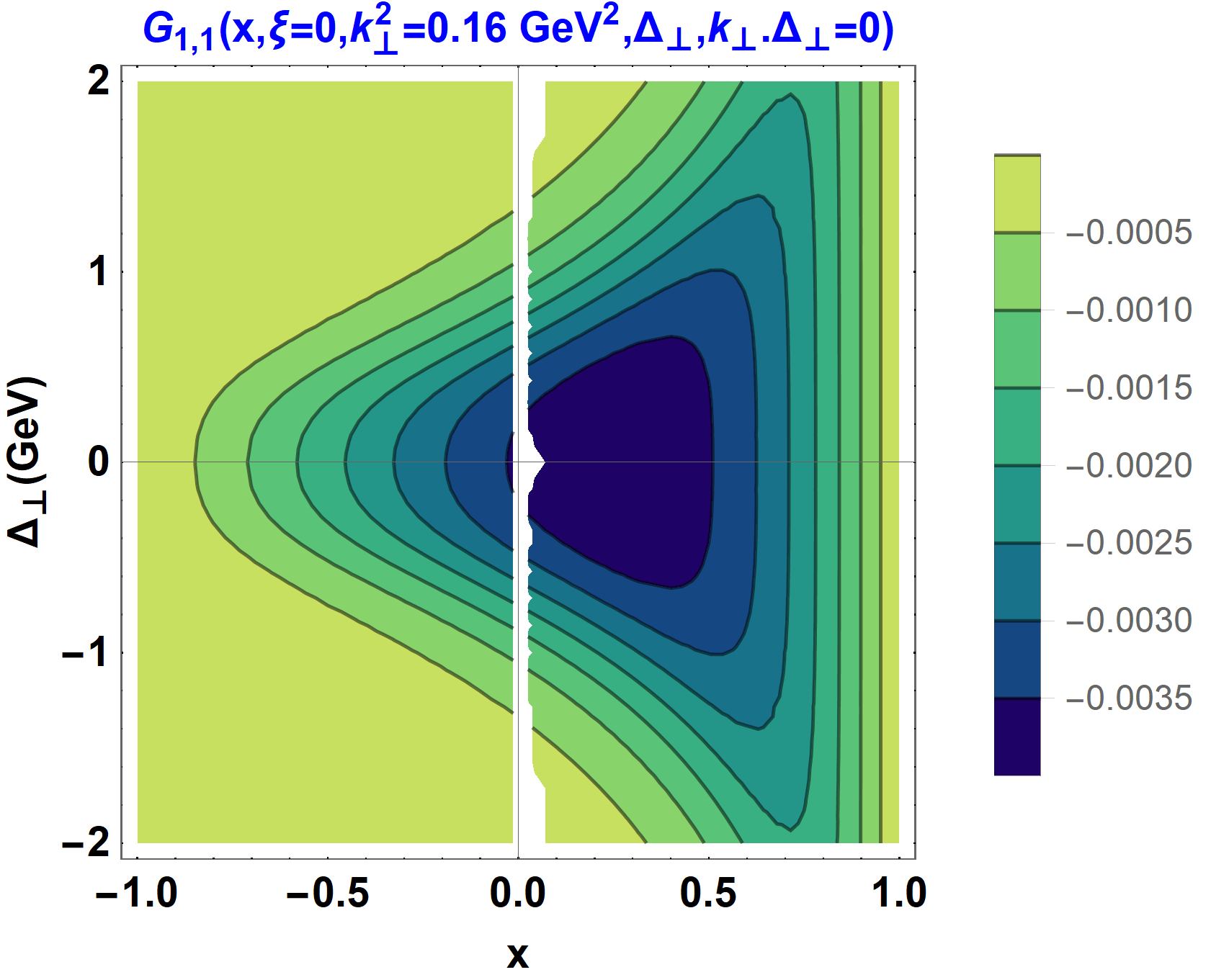}
\caption{$\rho_{B}=\rho_0$}
\end{subfigure}
\caption{\label{GTMDs_xdeltaa}Contour plots of the GTMDs $F_{1,4}$, $G_{1,4}$, and $G_{1,1}$ as functions of $x$ and $\Delta_\perp$, shown for baryonic densities $\rho_{B} = 0$ and $\rho_{B} = \rho_0$.}
\end{figure}
In this section, the behavior of the GTMDs $F_{14}$, $G_{11}$, and $G_{14}$ is examined as functions of the longitudinal momentum fraction $x$ and the transverse momentum transfer $\Delta_\perp$. We use Eq.~(\ref{AD4}) at $\rho_B=0$ and $\rho_B=\rho_0$ in the Eqs.~(\ref{F14})-(\ref{G14}) to generate the contour plots of GTMDs for eP and eA collisions environments, respectively, and they are presented in Fig.~\ref{GTMDs_xdeltaa}. In all plots, the skewness parameter is fixed at $\xi = 0$, and the quark transverse momentum is taken as $|k_\perp| = 0.4~\text{GeV}$. Furthermore, the configuration $k_\perp \perp \Delta_\perp$ is chosen so that the scalar product $k_\perp \cdot \Delta_\perp$ does not contribute to the expressions. It is observed that all distributions are symmetric about the axis $\Delta_\perp = 0$. This symmetry arises from the fact that the GTMDs depend only on $\Delta_\perp^2$. We also observe that a white strip appears in all the distributions due to the singular behavior at small $x$. This arises because all the distributions depend on the quantity $\alpha$, whose denominator contains a common factor of $x^2$. As $x \to 0$, this factor leads to a divergence in $\alpha$, resulting in the observed white strip in the numerical plots.

\begin{figure}[htp!]
\centering
\begin{subfigure}[b]{0.4\textwidth}
\centering
\includegraphics[width=\textwidth]{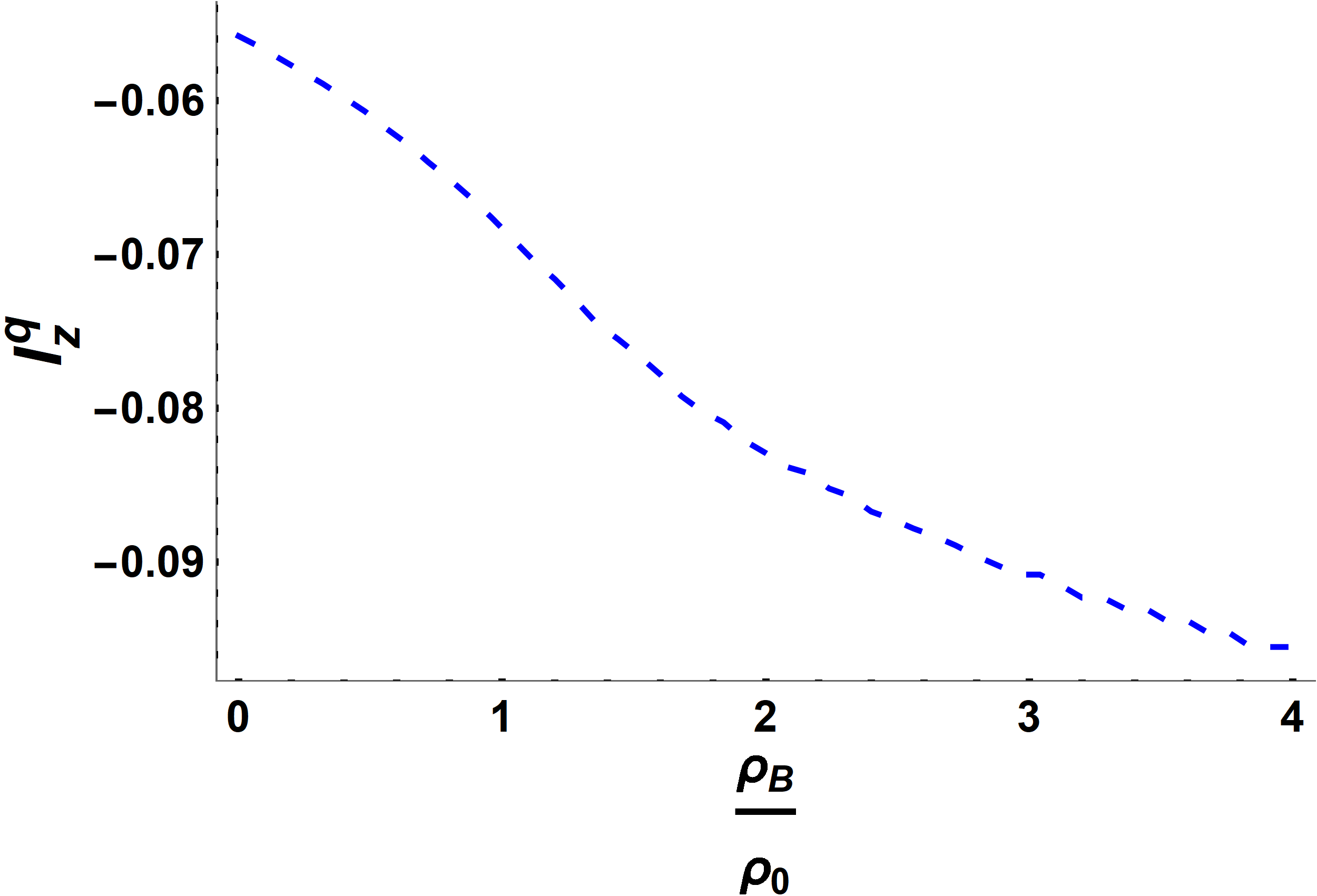}
\caption{}
\end{subfigure}
\centering
\begin{subfigure}[b]{0.4\textwidth}
\centering
\includegraphics[width=\textwidth]{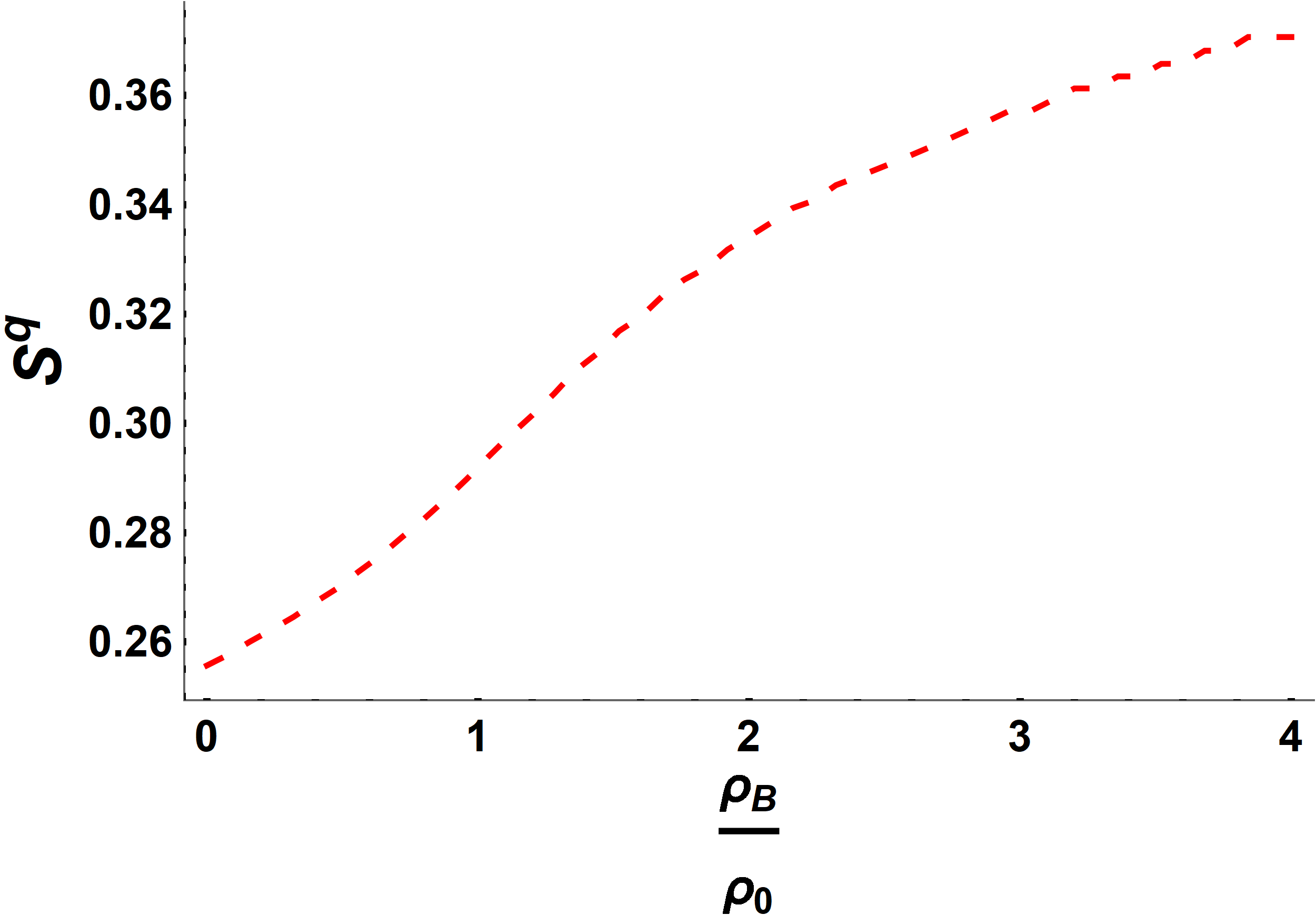}
\caption{}
\end{subfigure}
\hfill
\begin{subfigure}[b]{0.4\textwidth}
\centering
\includegraphics[width=\textwidth]{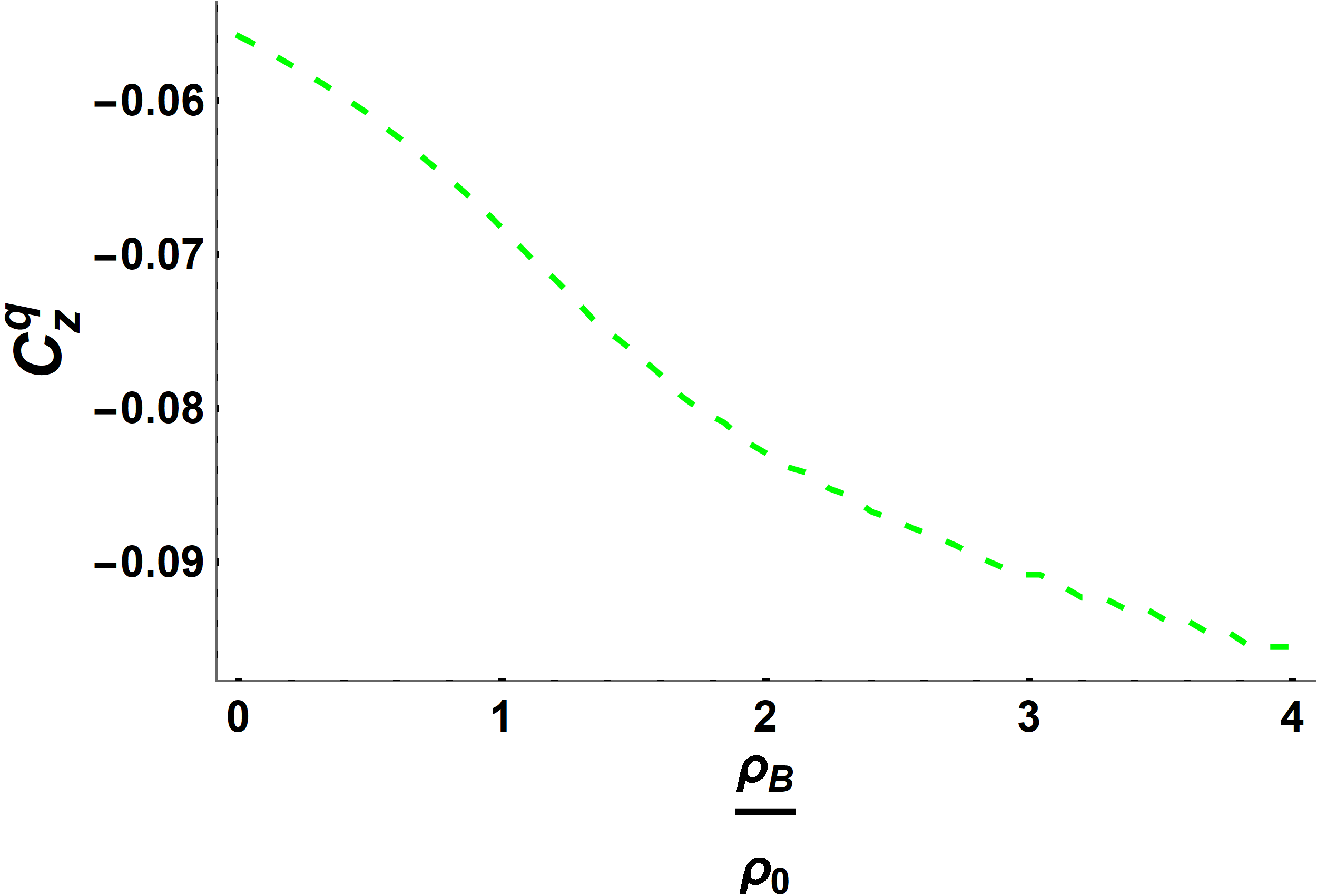}
\caption{}
\end{subfigure}
\caption{\label{OAM_density}Variation of (a) OAM, (b) spin, and (c) spin-orbit correlation as functions of normalized baryonic density $\rho_B/\rho_0$.}
\end{figure}

From Fig.~\ref{GTMDs_xdeltaa}(a) and (b), it is observed that the distribution $F_{1,4}$ attains its maximum in the positive $x$ region, while it is significantly suppressed for negative $x$. In addition, the distribution appears more localized at zero baryonic density ($\rho_{B} = 0$), whereas it becomes more spread out for finite density ($\rho_{B} = \rho_0$). One would naively think from the behavior in Fig.~\ref{GTMDs_xdeltaa} that in the presence of a nuclear medium, the magnitude of the OAM decreases as the magnitude of $F_{1,4}$ decreases. The inverse dependence of $l_{z}^{q}$ on $m^{\ast}(\rho_{B})$ reverses this trend (see Eq.~\ref{lqz}), and eventually one gets the quark OAM, which is comparatively higher than the vacuum. This is further analyzed in Fig.~\ref{OAM_density}(a). The negative sign in Eq.~\ref{lqz} basically indicates that OAM is anti-aligned with quark spin. 

From Fig.~\ref{GTMDs_xdeltaa}(c) and (d), the distribution $G_{14}$ is nonvanishing predominantly in the positive $x$ region. Furthermore, the magnitude of the distribution is larger at finite baryonic density ($\rho_{B} = \rho_0$) compared to the vacuum case ($\rho_{B} = 0$). Since $G_{14}$ is directly related to the quark spin contribution, this behavior indicates an enhancement of the quark spin component at normal nuclear matter density relative to the vacuum. 

From Fig.~\ref{GTMDs_xdeltaa}(e) and (f), the distribution $G_{11}$ is primarily concentrated in the positive $x$ region, with a comparatively smaller contribution in the negative $x$ domain. In addition, the magnitude of $G_{11}$ is suppresed at finite baryonic density ($\rho_{B} = \rho_0$) relative to the vacuum case ($\rho_{B} = 0$). Since $G_{11}$ is associated with the quark spin--orbit (L--S) correlation, this behavior indicates that the quark L--S coupling is weaker at normal nuclear matter density. 

In all GTMDs, the suppressed pattern in the negative $x$ region suggests a reduced contribution from antiquarks in this correlation. By grossly connecting these $\rho_B=0$ and $\rho_B=\rho_0$ results of GTMDs with eP and eA collision environments respectively, we can expect a modified the internal structure of the GTMDs $F_{14}$, $G_{14}$ and $G_{11}$ in terms of both magnitude and localization of the distribution, when one goes from eP to eA collision measurments.

Next, using Eqs.~(\ref{lqz}), (\ref{sq}) and (\ref{cqz}), we calculate quark OAM, spin, and spin-orbit (L-S) correlation, respectively, where constituent quark mass from Eq.~(\ref{AD4}) will be plugged in, which varies with density.
The results are presented in Fig.~\ref{OAM_density}, where panels (a), (b), and (c) correspond to OAM, spin, and spin--orbit correlation, respectively. Understanding the negative
sign of OAM as its anti-alignment with quark spin, if we consider only the magnitude of OAM, then an increasing trend with nuclear density is noticed in Fig.~\ref{OAM_density}(a). Similarly, quark spin and the magnitude of quark spin--orbit correlation also increase with
nuclear density. Reader can mark $\rho_B/\rho_0=0$ and 1 as proton and nucleus environment
of non-perturbative QCD domains and expect $40\%$, $16\%$ and $40\%$ relative enhancement of quark OAM, spin and L-S contribution. Here, we consider same upper limit of $k_\perp$ integration for proton and nucleus both cases. There is other possibility also by using appropriate density
dependent upper limit of $k_\perp$ integration, the quark OAM, spin and L-S contribution remain finally density independent. Future experimental measurement of quark OAM, spin and L-S contribution inside proton and nucleus via eP and eA collisions can only guide us about the methodology of integration. Therefore, instead of full integrated GTMDs (OAM, spin and L-S contribution), we can define a ratio between eA and eP collision measurement for GTMDs along x (longitudinal momentum fraction) axis, which is discussed in the next paragraph.

\begin{figure}[htp!]
\centering
\begin{subfigure}[b]{0.45\textwidth}
\centering
\includegraphics[width=\textwidth]{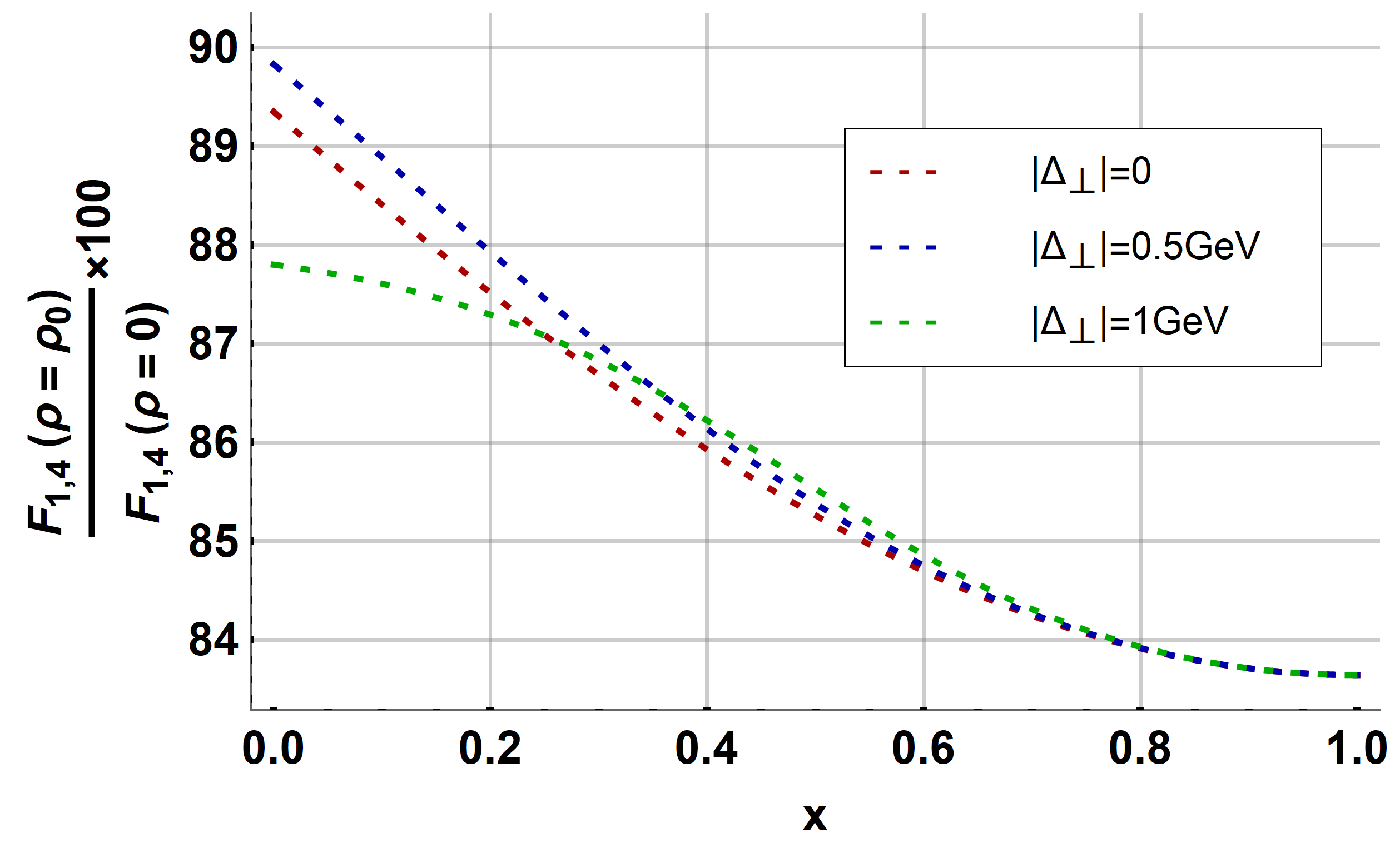}
\caption{}
\end{subfigure}
\centering
\begin{subfigure}[b]{0.45\textwidth}
\centering
\includegraphics[width=\textwidth]{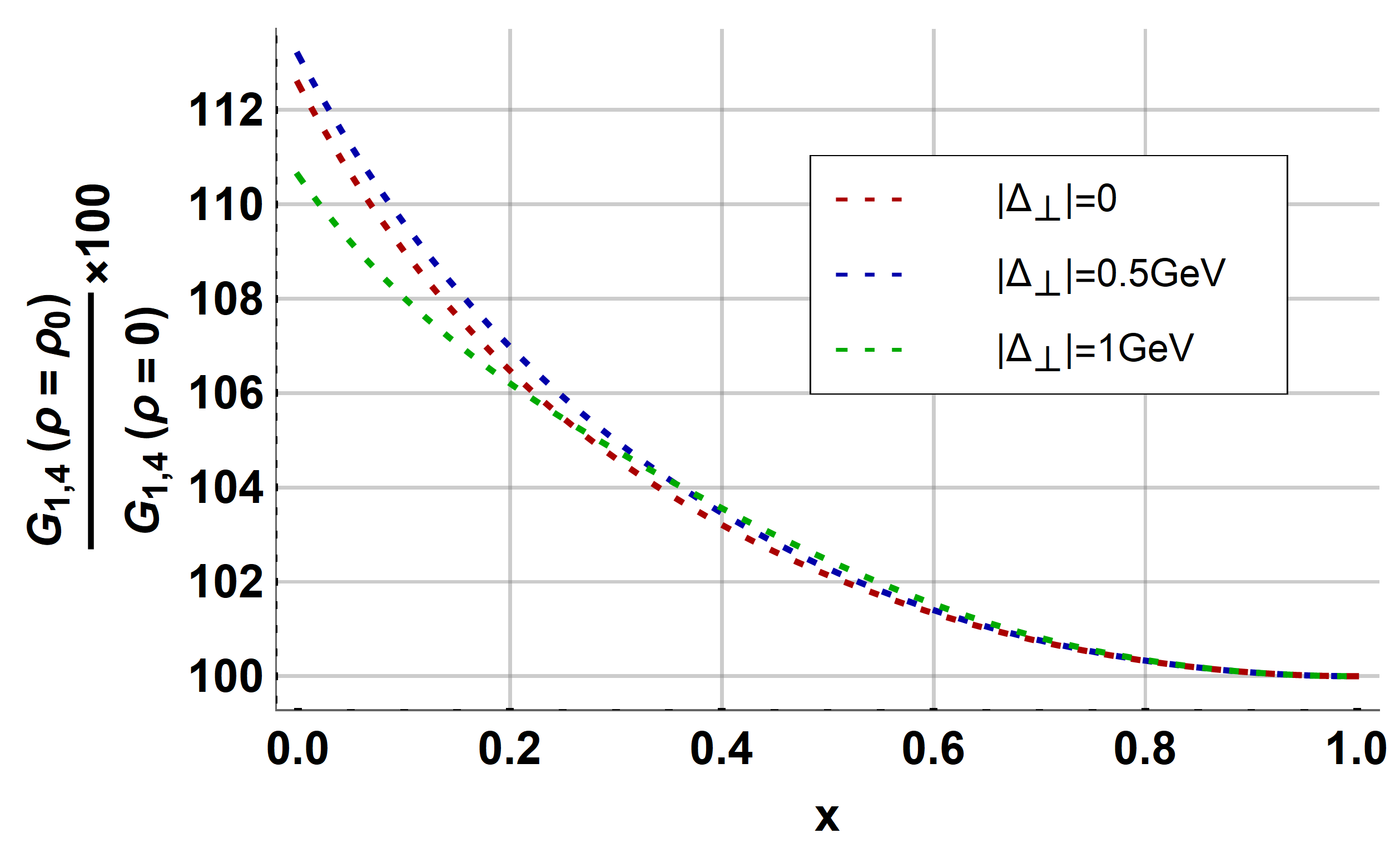}
\caption{}
\end{subfigure}
\hfill
\begin{subfigure}[b]{0.45\textwidth}
\centering
\includegraphics[width=\textwidth]{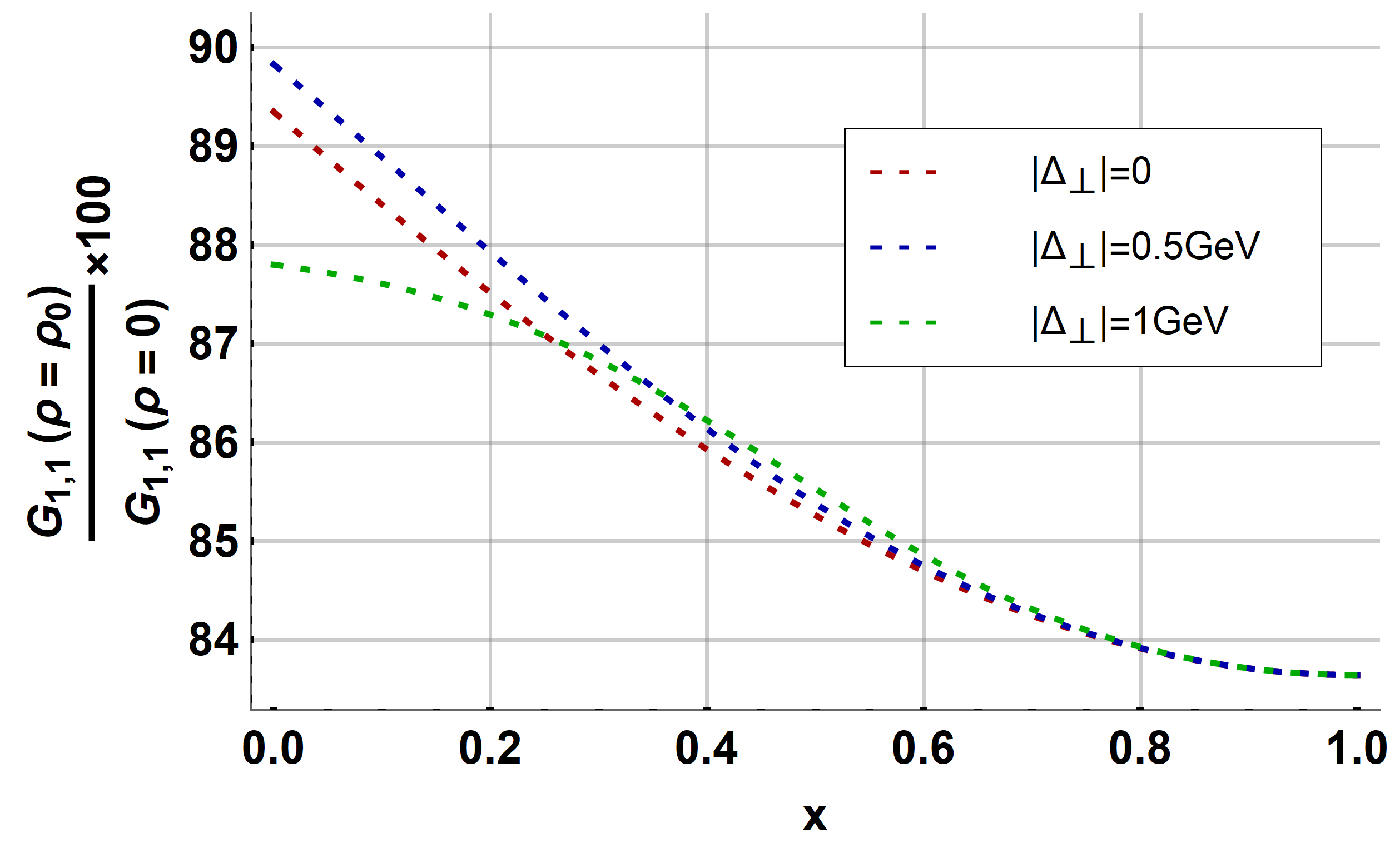}
\caption{}
\end{subfigure}
\caption{\label{GTMDs_error}Percentage deviation of the GTMDs (a) $F_{1,4}$, (b) $G_{1,4}$, and (c) $G_{1,1}$ as functions of $x$ for different values of $|\Delta_\perp|$.}
\end{figure}

Similar to the nuclear suppression quantity ($R_{AA}$) in heavy ion collision (HIC) experiments~\cite{ALICE:2012aqc}, we have defined a ratio between GTMDs at $\rho_B=\rho_0$ and $\rho_B=0$, whose deviation from 1 or 100 (in percentage scale) may be appeared as interesting quantities like ($R_{AA}$).
Figure~\ref{GTMDs_error} shows the percentage variation of the GTMDs $F_{14}$, $G_{14}$, and $G_{11}$ as functions of the longitudinal momentum fraction $x$ for different values of the transverse momentum transfer $\Delta_\perp$. It is observed that the percentage variation decreases monotonically with increasing $x$ for all three distributions.
At lower values of $x$, the variation is relatively larger, indicating a stronger sensitivity of the GTMDs to $\Delta_\perp$ in this region. As $x$ increases, the curves corresponding to different $\Delta_\perp$ values gradually converge, suggesting that the dependence on $\Delta_\perp$ becomes weaker at higher $x$.


The GTMD ratios associated with OAM and spin-orbit correlations inside the nucleus exhibit a suppression of approximately $10\%$--$16\%$ relative to their values inside a proton in vacuum, whereas the GTMD ratio associated with quark spin shows an enhancement of approximately $0\%$--$12\%$. To the best of our knowledge, this work provides the first theoretical prediction of these GTMD ratios in the dressed quark model, which could be measured at the future EIC at BNL through a comparison of eP and eA collisions. Analogous to the nuclear modification factor $R_{AA}$ in HICs, whose deviation from unity reflects the effects of the hot and dense QCD medium, the proposed GTMD ratios can serve as probes of nuclear density and many-body effects in the nonperturbative regime of QCD. In particular, deviations of these ratios from unity would provide a direct signature of such nuclear medium effects.

\section{Summary and Conclusion}\label{Sec:SC}
In summary, we have theoretically studied the distributions of quark spin and OAM within the proton and nuclei, which may be experimentally accessed in the future EIC facility at BNL, USA, where measurements in both eP and eA collisions are possible. For mapping the eP and eA collision environment, we have considered quark masses at zero baryon density and at nuclear saturation density, adopted from the NJL model framework. Using these two inputs of quark masses for mapping eP and eA collisions, respectively, we have calculated GTMDs within the light-front dressed quark model and provided explicit analytical expressions for the leading-twist distributions. We have focused only on those GTMDs that are connected to quark OAM, spin, and spin-orbit correlations.
Within this setup, we have graphically presented OAM, spin, and spin-orbit correlations as functions of density. The results indicate that the magnitude of quark OAM, spin-orbit correlation, and spin contribution increases with increasing baryon density.
We have proposed a possible phenomenological quantity, the ratio of GTMDs for eP and eA collisions, which can be measured in future experiments at the EIC facility. It is quite similar to $R_{AA}$ measurement in heavy ion collision experiments, which indirectly measure the QCD at finite temperature and density. Similarly, these GTMD ratios can be considered an indirect measure of the nuclear density effect in QCD.

\section{Acknowledgement}
This work was partially supported by the Ministry of Education (MoE), Government of India (A.D.), and by the Board of Research in Nuclear Sciences (BRNS) and the Department of Atomic Energy (DAE), Government of India, under Grant No. 57/14/01/2024-BRNS/313 (S.G.). The authors thank Tanmay Maji for the initial discussions during the WHEPP-2025 workshop, organized by IIT Hyderabad, where the idea for this work was first formulated. The authors thank Satyajit Puhan for fruitful discussions.
S.J. acknowledges the hospitality of IIT Bhilai during his visit, where the present work was initiated.


\bibliography{Ref_Quark.bib}
\bibliographystyle{unsrturl}
\end{document}